\documentclass[preprint,aps,prd,amsmath,amssymb]{revtex4}
\usepackage{graphicx}
\usepackage{color}
\newcommand{\beq}{\begin{eqnarray}}
\newcommand{\eeq}{\end{eqnarray}}

\newcommand{\bmp}{\noindent\begin{minipage}{16cm}}
\newcommand{\emp}{\end{minipage}\vskip 7mm} 

\usepackage{graphicx}
\usepackage{dcolumn}
\usepackage{bm}
\usepackage{amsmath}
\usepackage{amsfonts}
\usepackage{bbm}
\usepackage{subfigure}

\usepackage{ulem}

\newcommand{\drawsquare}[2]{\hbox{%
\rule{#2pt}{#1pt}\hskip-#2pt
\rule{#1pt}{#2pt}\hskip-#1pt
\rule[#1pt]{#1pt}{#2pt}}\rule[#1pt]{#2pt}{#2pt}\hskip-#2pt
\rule{#2pt}{#1pt}}
\newcommand{\Yfund}{\raisebox{-.5pt}{\drawsquare{6.5}{0.4}}}
\newcommand{\Ysymm}{\Yfund\hskip-0.4pt%
                    \Yfund}
\def\symm{\Ysymm}

\def\drawbox#1#2{\hrule height#2pt
        \hbox{\vrule width#2pt height#1pt \kern#1pt
              \vrule width#2pt}
              \hrule height#2pt}
\def\Fund#1#2{\vcenter{\vbox{\drawbox{#1}{#2}}}}
\def\Asym#1#2{\vcenter{\vbox{\drawbox{#1}{#2}
              \kern-#2pt 
              \drawbox{#1}{#2}}}}

\def\fund{\Fund{6.4}{0.3}}
\def\asymm{\Asym{6.4}{0.3}}

\begin{document}

\title{\hfill\vbox{\hbox{\rm\small CERN-PH-TH/2007-231}}\\{\Large Supersymmetry Inspired QCD Beta Function } }
\author{Thomas A. {\sc Ryttov}$^{a,c}$}
\email{ryttov@nbi.dk}
\author{Francesco {\sc Sannino}$^{b,c}$}
\email{sannino@ifk.sdu.dk} \affiliation{$^a$CERN Theory Division,
CH-1211 Geneva 23, Switzerland \\ $^b$University of Southern Denmark, Campusvej 55, DK-5230 Odense M \\
$^c$Niels Bohr Institute, Blegdamsvej 17, DK-2100 Copenhagen, Denmark}

\begin{abstract}
We propose an all orders beta function for ordinary Yang-Mills
theories with or without fermions inspired by the
Novikov-Shifman-Vainshtein-Zakharov beta function of ${\cal N}=1$
supersymmetric gauge theories. The beta function allows us to bound the conformal window. When restricting to one adjoint Weyl fermion we show how the proposed beta function matches the one of supersymmetric Yang-Mills theory. The running of the pure Yang-Mills coupling is computed and the
deviation from the two loop result is presented.  We then compare the deviation with the one obtained from lattice data also with respect to the two loop running.

\end{abstract}

\maketitle

\section{Introduction}
Inspired by the Novikov-Shifman-Vainshtein-Zakharov (NSVZ) supersymmetric beta function \cite{Novikov:1983uc,Shifman:1986zi,Jones:1983ip} we
 propose an all orders beta function for nonsupersymmetric gauge theories with or without massless
 fermions transforming according to arbitrary representations of the underlying  $SU(N)$ gauge group. The beta function
 at small coupling reduces to the  two loop beta function. The form of the  new beta function allows us to bound
 the phase diagram for a generic nonsupersymmetric gauge theory with fermionic matter in a given, but otherwise arbitrary,
 representation of the underlying gauge group. The result is simple and we compare it with the phase diagram presented
 in \cite{Sannino:2004qp,Dietrich:2006cm,Ryttov:2007sr} obtained using the truncated Schwinger-Dyson approximation also referred as ladder approximation in
 \cite{Appelquist:1988yc,Cohen:1988sq}. Further studies of the
nonsupersymmetric conformal window and its properties can be found
in
\cite{Appelquist:1996dq,MY,Sannino:1999qe,Harada:2003dc,Gies:2005as,Ndili:2005ni}.

We find that the ladder results provide a conformal
window systematically smaller than the one presented here. The
conformal windows we propose make use of the new beta function and
the condition of the absence of negative norm states in a conformal
field theory. The actual size of the conformal window may be smaller
than the one presented here which can be considered as a bound on
the size of the conformal window. In the supersymmetric case this
criterion provides, when extra checks can be performed
\cite{Seiberg:1994pq}, the actual size of the conformal window. The beta function is then generalized to the case of a gauge theory
with matter in different representations of the gauge group.

We consider the specific case of a single massless Weyl fermion in
the adjoint representation which corresponds to super Yang-Mills. By
directly comparing our expression with the super Yang-Mills result
\cite{Jones:1983ip,Novikov:1983uc,Shifman:1986zi} we determine the anomalous
dimension of an adjoint fermion.

At infinite number of colors a prediction of the beta function was made in \cite{Armoni:2003gp} for theories with matter in the two-index symmetric and antisymmetric representation of the gauge group and for Yang-Mills theory in \cite{Armoni:2003fb}. Our proposed supersymmetry-inspired beta function coincides with these results at infinite number of colors. Another attempt to produce the Yang-Mills beta function at large N was recently made in \cite{Bochicchio:2007za}. In this latter approach the beta function has a zero at a finite value of the coupling. If the zero signals a physical infrared fixed point it is reasonable to expect that the associated beta function does not correspond to pure Yang-Mills.  

The zero flavor limit, i.e. pure Yang-Mills, is a quite interesting
case since we can compare the running due to the new beta function
with lattice data for $SU(2)$, $SU(3)$ and $SU(4)$
\cite{Luscher:1992zx,Luscher:1993gh,Lucini:2007sa}. We find the new
beta function to compare well with data, capturing the fact that the
results do not depend on the number of colors when plotting the
running of the 't Hooft coupling. This result and the comparison
with data is rather encouraging.

We finally determine the ratio between the area of a given conformal
window to the associated  asymptotically free one and find that it
is universal, i.e. does not depend on the specific  matter
representation. A universal ratio was found earlier in the
supersymmetric case \cite{Ryttov:2007sr}. The ratio assumes the same
value in the supersymmetric and in the non-supersymmetric case. We
then generalize the phase diagram to the case of multiple matter
representations simultaneously affecting the gauge dynamics.
Following \cite{Ryttov:2007sr} we determine the size of the new
conformal regions and find a remarkably simple formula measuring the
ratio of conformal regions with respect to the associated
asymptotically free regions. Universality manifests again since the
ratios depend only on how many representations are considered but
not which ones.

A relevant application lies in the physics beyond the standard
model. {\it Minimal walking technicolor} theories
\cite{Sannino:2004qp,Dietrich:2006cm,Foadi:2007ue,Dietrich:2005jn}
are interesting examples models for dynamical breaking of the electroweak symmetry since they pass the electroweak precision tests. The walking dynamics was first introduced in
\cite{Holdom:1984sk,Holdom:1983kw,Eichten:1979ah,Holdom:1981rm,Yamawaki:1985zg,Appelquist:an,Lane:1989ej}.
By walking one refers to the fact that the underlying coupling
constant decreases much more slowly with the reference scale than in
the case of QCD-like theories. The very low number of flavors needed
to reach the conformal window, for certain representations, makes
the minimal walking theories amenable to lattice investigations.
Indeed, recent lattice results \cite{Catterall:2007yx} show that the theory
with two Dirac fermions in the adjoint representation of the $SU(2)$
gauge group possesses dynamic which is different from the one with
fermions in the fundamental representation. Our results may also be helpful in determining the physical spectrum of walking theories \cite{Appelquist:1998xf,Appelquist:1999dq} relevant for electroweak physics \cite{Foadi:2007ue}.

Yet, another interesting application of our work is as a study of the theoretical
landscape underlying the {\it unparticle} physics world proposed by Georgi
\cite{Georgi:2007ek,Georgi:2007si}.  CP and CPT properties
of unparticle physics have been studied in
\cite{Zwicky:2007vv}. The theories presented here, belonging to the various conformal
regions, are natural candidates for a {\it particle}
description of the unparticle world following
\cite{Fox:2007sy,Nakayama2007}.

\section{Introducing the NSVZ Supersymmetric Beta Function }

The gauge sector of a supersymmetric $SU(N)$ gauge theory consists
of a supersymmetric field strength belonging to the adjoint
representation of the gauge group. The supersymmetric field strength
describes the gluon and the gluino. The matter sector is taken to be
vectorial and to consist of $N_f$ chiral superfields $\Phi$ in the
representation $r$ of the gauge group and $N_f$ chiral superfields
$\tilde{\Phi}$ in the conjugate representation $\overline{r}$ of the
gauge group. The chiral superfield $\Phi$ (or $\tilde{\Phi}$)
contains a Weyl fermion and a complex scalar boson.

The generators $T_r^a,\, a=1\ldots N^2-1$ of the gauge group in the
representation $r$ are normalized according to
$\text{Tr}\left[T_r^aT_r^b \right] = T(r) \delta^{ab}$ while the
quadratic Casimir $C_2(r)$ is given by $T_r^aT_r^a = C_2(r)I$. The
trace normalization factor $T(r)$ and the quadratic Casimir are
connected via $C_2(r) d(r) = T(r) d(G)$ where $d(r)$ is the
dimension of the representation $r$. The adjoint
representation is denoted by $G$.

The exact beta function of supersymmetric QCD was first found in
\cite{Novikov:1983uc,Shifman:1986zi} and further investigated
in \cite{Arkani-Hamed:1997mj,Arkani-Hamed:1997ut}. For a given representation it takes the form
\begin{eqnarray}
\beta (g) &=& - \frac{g^3}{16\pi^2}
\frac{\beta_0+2T(r)N_f\gamma(g^2)}{1-\frac{g^2}{8\pi^2}C_2(G)} \ , \\
\gamma(g^2) &=& - \frac{g^2}{4\pi^2}C_2(r) + O(g^4) \ ,
\end{eqnarray}
where $g$ is the gauge coupling,  $\gamma (g^2) = -d \ln Z(\mu) /d \ln \mu $ is the anomalous
dimension of the matter superfield, $\mu$ the renormalization scale and $\beta_0 = 3C_2(G) - 2T(r)N_f$ is the first beta function coefficient.

For the reader's convenience in Table \ref{factors} we list the
explicit group factors for the representations used here. A complete
list of all of the group factors for any representation and the way
to compute them is available in Table II of \cite{Dietrich:2006cm}
and the associated appendix \footnote{The normalization for the
generators here is different than the one adopted in
\cite{Dietrich:2006cm}.}.

\begin{table}
\begin{center}
    \begin{tabular}{c||ccc }
    r & $ \quad T(r) $ & $\quad C_2(r) $ & $\quad
d(r) $  \\
    \hline \hline
    $ \fund $ & $\quad \frac{1}{2}$ & $\quad\frac{N^2-1}{2N}$ &\quad
     $N$  \\
        $\text{$G$}$ &\quad $N$ &\quad $N$ &\quad
$N^2-1$  \\
        $\symm$ & $\quad\frac{N+2}{2}$ &
$\quad\frac{(N-1)(N+2)}{N}$
    &\quad$\frac{N(N+1)}{2}$    \\
        $\asymm$ & $\quad\frac{N-2}{2}$ &
    $\quad\frac{(N+1)(N-2)}{N}$ & $\quad\frac{N(N-1)}{2}$
    \end{tabular}
    \end{center}
\caption{Relevant group factors for the representations used
throughout this paper. However, a complete list of all the group
factors for any representation and the way to compute them is
available in Table II and the appendix of
\cite{Dietrich:2006cm}.}\label{factors}
    \end{table}
\section{NSVZ - Inspired Non Supersymmetric Beta Function}
Consider now a generic non supersymmetric gauge theory with $N_f $ Dirac fermions in a given representation $r$
of the gauge group.
The beta function  to two loops reads:
\begin{eqnarray}\beta (g) = -\frac{\beta_0}{(4\pi)^2} g^3 - \frac{\beta_1}{(4\pi)^4} g^5 \ ,
\label{perturbative}
\end{eqnarray}
where $g$ is the gauge coupling and the beta function coefficients are given by
\begin{eqnarray}
\beta_0 &=&\frac{11}{3}C_2(G)- \frac{4}{3}T(r)N_f \\
\beta_1 &=&\frac{34}{3} C_2^2(G)
- \frac{20}{3}C_2(G)T(r) N_f  - 4C_2(r) T(r) N_f  \ .\end{eqnarray}
To this order the two coefficients are universal,
i.e. do not depend on which renormalization group scheme one has used to determine them.
The perturbative expression for the anomalous dimension reads:
\begin{equation}
\gamma(g^2) = \frac{3}{2} C_2(r) \frac{g^2}{4\pi^2} + O(g^4) \ .
\end{equation}
With $\gamma =-{d\ln m}/{d\ln \mu}$ and $m$ the renormalized fermion mass.
It would be great to have the complete expression for the beta
function for a non supersymmetric theory. This seems to be a
formidable task. Inspired by supersymmetry we suggest an all orders
non supersymmetric beta function which has a number of interesting
properties and predictions which we will compare and test against
nonperturbative results found using various methods or models.

The first observation is that the perturbative anomalous
dimension depends on $C_2(r)$ which
appears explicitly in the last term of the second coefficient of the beta function.
We hence write the beta
function in the following form:
\begin{eqnarray}
\beta(g) &=&- \frac{g^3}{(4\pi)^2} \frac{\beta_0 - \frac{2}{3}\, T(r)\,N_f \,
\gamma(g^2)}{1- \frac{g^2}{8\pi^2} C_2(G)\left( 1+ \frac{2\beta_0'}{\beta_0} \right)} \ ,
\end{eqnarray}
with
\begin{eqnarray}
\beta_0' &=& C_2(G) - T(r)N_f  \ .
\end{eqnarray}
It is a simple matter to show that the above beta function reduces
to Eq. (\ref{perturbative}) when expanding to $O(g^5)$. Given that
only the two loop beta function has universal coefficients, i.e. is
independent of the renormalization scheme, we will assume the
existence of a scheme for which our beta function is complete.
\section{IR Fixed Point}
As we decrease the number of flavors from just below the point where asymptotic freedom is lost, corresponding to:
\begin{eqnarray}
N_f^{\rm{I}} = \frac{11}{4} \frac{C_2(G)}{T(r)} \ ,
\end{eqnarray}
one expects a perturbative zero in the beta function to occur \cite{Banks:1981nn}. From the expression proposed above one finds that at the zero of the beta function, barring zeros in the denominator,  one must have
\begin{eqnarray}
\gamma = \frac{11C_2(G)-4T(r)N_f}{2T(r)N_f} \ .\end{eqnarray}
The anomalous dimension at the IR fixed point is small for a value of $N_f$ such that:
\begin{equation}
N_f = N_f^I (1 - \epsilon) \ , \quad {\rm with} \quad {\epsilon > 0} \ ,
\end{equation}
and $\epsilon \ll 1$. Indeed, in this approximation we find:
\begin{equation}
\gamma = \frac{2\epsilon}{1-\epsilon} \ll 1 \ .
\end{equation}
It is also clear that the value of $\gamma$ increases as we keep decreasing the number of flavors.
Before proceeding let us also analyze in more detail the denominator of our beta function.
At the infrared fixed point we have:
\begin{equation}
1- \frac{g^2_{\ast}}{8\pi^2} \, C_2(G)\,\frac{1}{2}\,\left(5 - \frac{21}{11\epsilon}  \right) \ ,
\end{equation}
For very small $\epsilon$ the denominator is positive while staying
finite as $\epsilon$ approaches zero. The finiteness of the denominator
is due to the fact that from the perturbative expression of the
anomalous dimension (valid for small epsilon) the fixed point value
of $g_{\ast}$ is:
\begin{eqnarray}
\frac{g^2_{\ast}}{8\pi^2} = \epsilon \,\,\frac{2}{3C_2(r)} + O(\epsilon^2) \ .
\end{eqnarray}

\begin{figure}[h]
\resizebox{10cm}{!}{\includegraphics{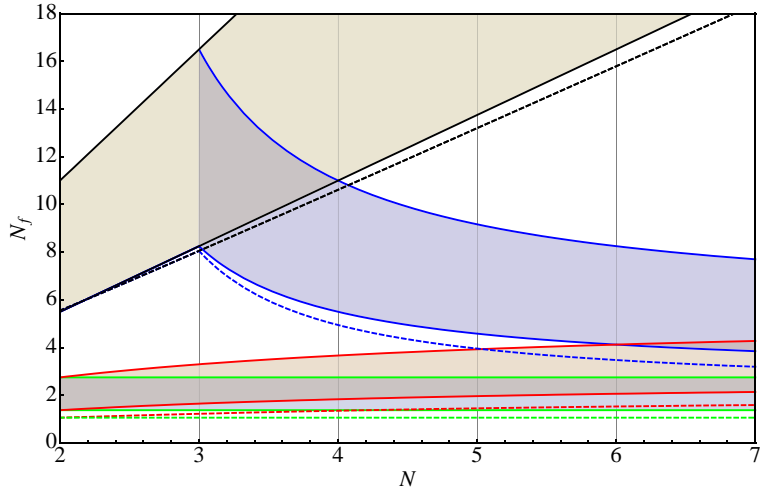}}
\caption{Phase diagram for nonsupersymmetric theories with fermions
in the: i) fundamental representation (black), ii) two-index
antisymmetric representation (blue), iii) two-index symmetric
representation (red), iv) adjoint representation (green) as a
function of the number of flavors and the number of colors. The
shaded areas depict the corresponding conformal windows. Above the
upper solid curve  the theories are no longer asymptotically free.
Between the upper and the lower solid curves the theories are
expected to develop an infrared fixed point according to the NSVZ
inspired beta function. The dashed curve represents the change of
sign in the second coefficient of the beta function.} \label{PHNew}
\end{figure}

Since a perturbative fixed point does exist we extend the analysis to a lower number of flavors. The dimension of the chiral condensate is $D(\bar{\psi} \psi)=3-\gamma$ which at the IR fixed point value reads
\begin{equation}
D (\bar{\psi} \psi)= \frac{10T(r)N_f - 11C_2(G)}{2T(r)N_f} \ .
\end{equation}
 To avoid negative norm states in a conformal field theory one must have $D\geq 1$ for non-trivial
 spinless operators \cite{Mack:1975je,Flato:1983te,Dobrev:1985qv}.
 Hence the critical number of flavors below which the unitarity bound is violated
 according to the NSVZ inspired beta function is
\begin{eqnarray}
N_f^{\rm{II}} = \frac{11}{8} \frac{C_2(G)}{T(r)} \ ,
\end{eqnarray}
which corresponds to having set $\gamma=2$. One should note that the analysis above is
similar to the one done for supersymmetric gauge theories \cite{Seiberg:1994pq}.
However, the actual size of the conformal window may be smaller than the one
presented here which hence can be considered as a bound on the size of the window. In Figure \ref{PHNew} we plot the new phase diagram. Our conformal bound in the case of fermions transforming according to the fundamental representation of the $SU(3)$ gauge group predicts that a physical infrared fixed point can be reached, in this case, for a number of Dirac flavors larger than 8.25 which is larger than the one found by Iwasaki et al. \cite{Iwasaki:2003de} which is around six.  

\subsection{Comparison with the Ladder approximation}

 We now confront our bound of the conformal window with the one obtained using the ladder approximation \cite{Dietrich:2006cm}. In the ladder approximation one finds:
\begin{eqnarray}
{N_f^\mathrm{II}}_{\rm Ladder} &=& \frac{17C_2(G)+66C_2(r)}{10C_2(G)+30C_2(r)}
\frac{C_2(G)}{T(r)} \ . \label{nonsusy}
\end{eqnarray}
This value is very crude \cite{Appelquist:1988yc,Cohen:1988sq}. Comparing with the NSVZ inspired result we see that it is the coefficient of $C_2(G)/T(r)$ which is different.

\begin{figure}[h]
\resizebox{10cm}{!}{\includegraphics{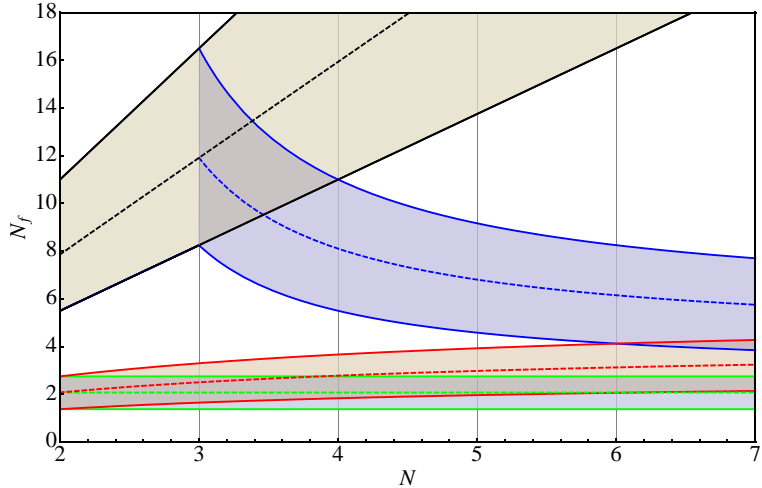}}
\caption{Phase diagram for nonsupersymmetric theories with fermions
in the: i) fundamental representation (black), ii) two-index
antisymmetric representation (blue), iii) two-index symmetric
representation (red), iv) adjoint representation (green) as a
function of the number of flavors and the number of colors. The
shaded areas depict the corresponding conformal windows. Above the
upper solid curve  the theories are no longer asymptotically free.
In between the upper and the lower solid curves the theories are
expected to develop an infrared fixed point according to the NSVZ
inspired beta function. The area between the upper solid curve and
the dashed curve corresponds to the conformal window obtained in the
ladder approximation.} \label{PHComparison}
\end{figure}

To better appreciate the differences between these two results we plot the two
conformal windows predicted within these two methods in Figure
\ref{PHComparison} for four types of fermion representation.

The ladder result provides a size of the window, for every  fermion representation, smaller than the bound found with our approach. This is a consequence of the value of the anomalous dimension at the lower bound of the window. The unitarity constraint corresponds to $\gamma =2$ while the ladder result is closer to $\gamma \sim 1$. Indeed if we pick $\gamma =1$ our conformal window approaches the ladder result. Incidentally, a value of $\gamma$ larger than one, still allowed by unitarity, is a welcomed feature when using this window to construct walking technicolor theories. It allows for the physical value of the mass of the top while avoiding a large violation of flavor changing neutral currents \cite{Luty:2004ye} which were investigated in  \cite{Evans:2005pu} in the case of the ladder approximation for minimal walking models.

\subsection{Generalization to Multiple Representations}
The generalization for a generic gauge theory with massless fermions in $k$ different representations is:
\begin{eqnarray}
\beta(g) &=&- \frac{g^3}{(4\pi)^2} \frac{\beta_0 - \frac{2}{3}\, \sum_{i=1}^k T(r_i)\,N_{f}(r_i) \,\gamma_i}{1- \frac{g^2}{8\pi^2} C_2(G)\left( 1+ \frac{2\beta_0'}{\beta_0} \right)} \ ,
\end{eqnarray}
with
\begin{eqnarray}
\beta_0' &=& C_2(G) - \sum_{i=1}^k T(r_i)N_f(r_i)  \ ,
\end{eqnarray}
and
\begin{eqnarray}
\beta_0 &=&\frac{11}{3}C_2(G)- \frac{4}{3}\sum_{i=1}^k \,T(r_i)N_f(r_i)  \ .
\end{eqnarray}

\section{Matching to Exact Results and Lattice Data}
We now take different limits in theory space and, in doing so, we will gain confidence on the validity of the NVSZ -inspired beta function. We first recall how to relate the gauge singlet bilinear fermion condensate at different energy scales in the case of the canonically normalized fermion kinetic term $\bar{\psi}\gamma^{\mu}D_{\mu} \psi$:
\begin{equation}
\langle \bar{\psi}\psi\rangle _{Q}  =  \exp\left[\int_{\mu}^{Q} dg\, \frac{\gamma(g)}{\beta(g)}\right]  \,\langle \bar{\psi} \psi\rangle_{\mu} \ .
\end{equation}
Here $\bar{\psi} \psi$ is a gauge singlet operator and we have suppressed the color and flavor indices.
At the lowest order in perturbation theory one obtains the simple formula:
\begin{equation}
\langle \bar{\psi}\psi\rangle _{Q}  =  \left[ \frac{g(\mu)^2}{g(Q)^2}\right]^{\frac{3C_2(r)}{\beta_0}} \,\langle \bar{\psi} \psi\rangle_{\mu} \ ,
\end{equation}
with $r$ the representation of the Dirac fermion $\psi$. By construction and at the lowest order in perturbation theory the operator
\begin{equation}
\left[g(Q)^2\right]^{\frac{3C_2(r)}{\beta_0}}\langle\bar{\psi} \psi \rangle _{Q} \ ,
\end{equation} is renormalization group invariant.

\subsection{Super Yang-Mills}
Consider the theory with one single Weyl fermion transforming according to the adjoint representation of the gauge group. The beta function reads:
\begin{equation}
\beta(g) = -\frac{g^3}{(4\pi)^2}3N\frac{1-\frac{\gamma_{\rm Adj}}{9}}{1-\frac{g^2}{8\pi^2}\frac{4N}{3}} \ ,
\label{SYM-Inspired}
\end{equation}
with $\gamma_{\rm Adj}$ the anomalous dimension of the fermion condensate. This theory corresponds to super Yang-Mills for which we know the result \cite{Jones:1983ip,Novikov:1983uc}:
\begin{equation}
\beta_{SYM}(g) = -\frac{g^3}{(4\pi)^2}\frac{3N}{1-\frac{g^2}{8\pi^2}N} \ .
\end{equation}
In the NSVZ expression above there is no explicit appearance of the anomalous dimension while this is manifest in Eq.~(\ref{SYM-Inspired}). The absence of the anomalous dimension in the NSVZ form of the beta function is due to the choice of normalization of the gluino condensate which renders the associated operator renormalization group invariant. Assuming that the two beta functions have been computed in the same renormalization scheme we can equate them. This provides the expression for the anomalous dimension of the fermion bilinear in the adjoint representation of the gauge group normalized in the standard way:
\begin{equation}
\gamma_{\rm {Adj}} = \frac{g^2}{8\pi^2} \frac{3N}{1-\frac{g^2}{8\pi^2} N} \ .
\end{equation}
Note that in our scheme we have that
\begin{equation}
g^2(Q) \langle\lambda \lambda\rangle_{Q} \ ,
\end{equation}
is a renormalization group invariant quantity to all orders. This is exactly the definition of the gaugino condensate used by NSVZ. One should also note that we recover the perturbative expression of $\gamma_{Adj}$ when expanding to $O(g^2)$.
\subsection{Pure Yang-Mills and Comparison with Lattice Data}
Pure Yang-Mills is an excellent study case, since it has been widely investigated in the literature, and much is known, especially via lattice simulations. Setting the number of flavors to zero we have:
\begin{equation}
\beta_{YM}(g)= -\frac{g^3}{(4\pi)^2}\frac{{\beta}_0}{1-\frac{g^2}{(4\pi)^2}\frac{{\beta}_1}{{\beta}_0}}\ ,
\end{equation}
with
\begin{equation}
\beta_0=\frac{11N}{3} \ , \qquad \beta_1=\frac{34N^2}{3} \ ,
\end{equation}
respectively for the one and two loop coefficients of the beta function. These are the only universal coefficients of a generic beta function in any scheme.
We now integrate the above beta function and compare our running coupling constant with the two loop result and find:
\begin{eqnarray}
\mu = \Lambda_{1}\exp\left[\frac{8\pi^2}{g^2\beta_0}\right]\left( g^2\beta_0\right)^\frac{\beta_1}{2\beta_0^2}\ , \qquad {\rm NSVZ -inspired}\end{eqnarray}
to be compared with the two loop beta function result:
\begin{eqnarray}
\mu = \Lambda_{2}\exp\left[\frac{8\pi^2}{g^2\beta_0}\right]\left( g^2\beta_0\right)^\frac{\beta_1}{2\beta_0^2}\left( 1 +\frac{ g^2}{16\pi^2}\frac{\beta_1}{\beta_0}\right)^\frac{-\beta_1}{2\beta_0^2} \ .\qquad {\rm 2~loops}\end{eqnarray}
Note that we have normalized the invariant scales $\Lambda_i$ in such a way that they do not depend on the number of colors. It is also clear that the two results do not depend on the number of colors when considering $g^2N$ as the coupling, i.e. the 't Hooft coupling.

\begin{figure}[h]
\resizebox{10cm}{!}{\includegraphics{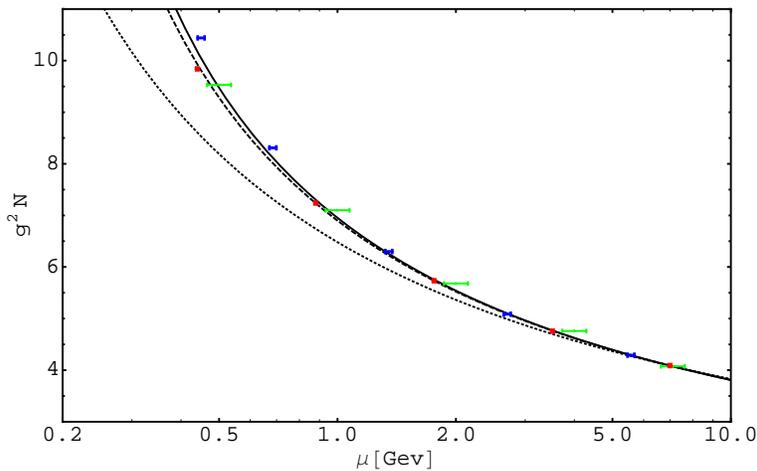}}
\caption{The evolution of the gauge coupling squared times the number of colors (i.e. the 't Hooft coupling) as a function of the energy scale for two, three and four colors. The solid curve is obtained using the susy inspired beta function, the dashed is obtained via the two loop beta function while the dotted curve is the one loop result. The green dots (biggest errorbars) correspond to lattice data for $SU(2)$ \cite{Luscher:1992zx}, the blue dots to $SU(3)$ \cite{Luscher:1993gh} and the red dots (smallest errorbars) to $SU(4)$ \cite{Lucini:2007sa}.}
\label{SU(n)}
\end{figure}

It is instructive to compare the deviation from the two loop result of the NSVZ-inspired beta function with the deviation of the lattice data also with respect to the two loop one.
In Figure \ref{SU(n)} we show the evolution of the 't Hooft coupling as a function of the energy scale and plot it together with the two, three and four colors  lattice data.

The solid curve is obtained using the NSVZ-inspired beta function, the dashed is obtained via the two loop beta function while the dotted curve is the one loop result. The green dots (biggest errorbars) correspond to lattice data for $SU(2)$ taken from \cite{Luscher:1992zx}, the blue dots to $SU(3)$ \cite{Luscher:1993gh} and the red dots (smallest errorbars) to $SU(4)$ \cite{Lucini:2007sa}. Despite the fact that the two renormalization schemes are different the size of the corrections with respect to the two loop coming from the lattice data and  the present beta function are similar. We find this result encouraging.

\section{Re-sizing the unparticle world: A New Universal Ratio}
Georgi has recently proposed to couple a conformal sector to the
Standard Model \cite{Georgi:2007ek}. In \cite{Ryttov:2007sr} we suggested a measure of how large, in theory space, the fraction of the
unparticle world is. We assumed, following Georgi, the unparticle
sector to be described, at the underlying level, by asymptotically
free gauge theories developing an infrared fixed point. We showed that a reasonable
measure is then, for a given representation, the ratio of the area of the
conformal window to that of the total window for asymptotically free gauge
theories
\begin{equation}
R_{FP} = \frac{\int_{N_{min}}^{\infty} N_f^{\mathrm I} \,dN
-\int_{N_{min}}^{\infty}  N_f^{\mathrm {II}} \,
dN}{\int_{N_{min}}^{\infty} N_f^{\mathrm I}\, dN }  \ ,
\end{equation}
where $N_{min}$ is the smallest number of colors permitted for
the chosen representation.

Remarkably, the above ratio turned out to be universal, i.e. independent of the matter representation, for any ${\cal N}=1$ supersymmetric theory. The value being 1/2. Using the NSVZ inspired beta function for nonsupersymmetric theories we again find the same universal result:
\begin{eqnarray}
R_{FP} = \frac{\frac{11}{4}-\frac{11}{8}}{\frac{11}{4}} = \frac{1}{2} \ . \qquad \qquad {\rm NSVZ - inspired}
\end{eqnarray}

A generic gauge theory will, in general, have matter transforming
according to distinct representations of the gauge group. We follow
the analysis first performed in  \cite{Ryttov:2007sr} of the conformal region for a generic $SU(N)$
gauge theory with ${N_f}(r_i)$ vector-like matter fields
transforming according to the representation $r_i$ with
$i=1,\ldots,k$ . We shall consider the non-supersymmetric case here and will use the NSVZ-inspired beta function to determine the fraction of conformal regions.

The generalization to $k$ different representations for
the expression determining the region in flavor space above which
asymptotic freedom is lost is simply
\begin{eqnarray}
\sum_{i=1}^{k}\frac{4}{11}T(r_i)N_f(r_i) = C_2(G) \ .
\end{eqnarray}

Following \cite{Ryttov:2007sr} we estimate the region above which the theories
develop an infrared fixed point via the following expression
\begin{eqnarray}
\sum_{i=1}^{k} \frac{8}{11} T(r_i) N_f(r_i) = C_2(G) \ ,
\end{eqnarray}

The volume, in flavor and color space,
occupied by a generic $SU(N)$ gauge theory is defined to be:
\begin{eqnarray}
V_{\eta}[N_{min}, N_{max}] &=& \int_{N_{min}}^{N_{max}} dN
\prod_{i=1}^{k} \int_{0}^{\frac{C_2(G) - \sum_{j=2}^{i} \eta
T(r_j)N_f(r_j)}{\eta T(r_{i+1})}} N_f(r_{i+1}) \ ,
\end{eqnarray}
with $\eta$ reducing to the number $4/11$ when the region to be evaluated is associated to the asymptotically free one and to $8/11$ when the region is the one below which one does not expect the occurrence of an infrared fixed point. The notation is such that $T(r_{k+1})\equiv
T(r_1)$, $N_f(r_{k+1}) \equiv N_f(r_1)$ and the sum
$\sum_{j=2}^{i}\eta \,T(r_j)N_f(r_j)$ in the upper limit of the flavor
integration vanishes for $i=1$. We defined the volume within a
fixed range of number of colors $N_{min}$ and $N_{{max}}$.

Hence the fraction of the conformal region to the region occupied by
the asymptotically free theories is, for a given number of
representations $k$:
\begin{eqnarray}
R_{FP} = \frac{V_{\frac{4}{11}}[N_{min}, N_{max}] -
V_{\frac{8}{11}}[N_{ {min}},N_{
{max}}]}{V_{\frac{4}{11}}[N_{ {min}},N_{ {max}}]} = 1-\left(\frac{1}{2}\right)^k\ .
\end{eqnarray}
Quite surprisingly the result obtained using the NSVZ-inspired beta function does not depend on which representation one uses but depends solely on the number $k$ of representations present. We recover $1/2$ for $k=1$. We estimated this ratio in the case of the ladder approximation first in \cite{Ryttov:2007sr}. We noticed then a small dependence which, however, can be related to the large uncertainty stemming from the ladder approximation. The bound of the conformal region is larger than the one computed using the ladder approximation.

\section{Conclusions}

We suggested an all orders beta function for ordinary Yang-Mills theories with or without  fermions inspired by the NSVZ beta function of ${\cal N}=1$ super gauge theories. We computed the bound on the conformal regions and then showed how the proposed beta function can be matched to the NSVZ one for super Yang-Mills.

By setting the number of matter flavors to zero one has the Yang-Mills beta function for any number of colors. Interestingly the latter depends only on the first two numerical coefficients of the beta function which are universal according to 't Hooft. The running of the 't Hooft coupling was computed and the
deviation from the two loop result presented.  We then compared this with the deviation of the lattice data also with respect to the two loop running. We found the size of the two deviations to be rather close.

Finally we examined the ratio between the area of conformal windows to the asymptotically free ones and showed that it is universal, i.e. does not depend on the specific  matter representation. A universal ratio was found earlier in the supersymmetric case \cite{Ryttov:2007sr}.

\acknowledgments
We gladly thank D. Anselmi, A. Armoni, M. Bochicchio, L. Del Debbio,  D.D. Dietrich, P. Di Vecchia, R. Foadi, M. T. Frandsen, A. Jarosz, D.R.T. Jones, M. J\"arvinen, B. Lucini, C. N\'u\~nez, K. Tuominen and R. Zwicky for discussions and/or careful reading of the manuscript. The work of T.R is supported by a Marie Curie Early Stage Research Training Fellowship of the European Community's Sixth Framework Programme under contract number MEST-CT-2005-020238-EUROTHEPHY. The work of F.S. is supported by the Marie Curie Excellence Grant under contract MEXT-CT-2004-013510.

\end{document}